\documentclass[%
 aip,
cp,  
 amsmath,amssymb,
  reprint,%
]{revtex4-2}

\usepackage{graphicx}
\usepackage{dcolumn}
\usepackage{bm}
\usepackage[utf8]{inputenc}
\usepackage[T1]{fontenc}
\usepackage{mathptmx} 

\begin{document}

\title{Results of the \textsc{Majorana Demonstrator's} Search for Neutrinoless Double Beta Decay}

\author{David J. Tedeschi} 
 \email[Corresponding author: ]{djtedesc@mailbox.sc.edu}
\affiliation{
  For the \textsc{Majorana} Collaboration \\
  University of South Carolina \\
  Department of Physics and Astronomy \\
  712 Main St., Columbia SC 29208
}

\date{\today}

\begin{abstract}
The \textsc{Majorana Demonstrator} recently concluded its search for neutrinoless double-beta decay. The experiment operated an array of up to 40.4 kg of germanium detectors, 27 kg of which were isotopically enriched in $^{76}$Ge and housed inside a compact shield consisting of lead and copper at the Sanford Underground Research Facility (SURF) in Lead, SD. The experiment achieved a world leading energy resolution of 0.12\% FWHM at 2039 keV, one of the lowest background rates in the region of the 0$\nu\beta\beta$ Q-value, 15.7 $^{+1.4}_{-1.3}$ cnts/(FWHM t y), and set a 0$\nu\beta\beta$ half-life limit of T$_{1/2}$ > 8.3×10$^{25}$ yrs based on its full active exposure of 65 kg yr, resulting in limits on the effective neutrino mass of m$_{\beta \beta}$ < 113-269 meV.
\end{abstract}

\maketitle

\section{\label{sec:Intro} Introduction}

In two-neutrino double-beta decay
(2$\nu\beta\beta$), the nucleus emits 2 $\beta$ particles and 2 $\overline{\nu } _{e}$ conserving lepton number.
It is an allowed second-order weak process that occurs in nature, although its rate is extremely low. Half-lives for this
decay mode have been measured at $\sim$10$^{19}$ years or longer in several nuclei.\cite{RevModPhys.95.025002}
Neutrinoless double-beta decay, 0$\nu\beta\beta$, where only the 2 $\beta$ particles are emitted with no neutrino, violates lepton number conservation 
and hence requires physics beyond the Standard Model. One can visualize $0\nu\beta\beta$ as an exchange
of a virtual neutrino between two neutrons within the nucleus. In the framework of the 
SU$_{L}(2) \times$ U$(1)$ Standard Model of weak interactions, the first neutron emits a right-handed anti-neutrino. However, the second
neutron requires the absorption of a left-handed neutrino. In order for this to happen, the neutrino must have mass so that it is not in
a pure helicity state and the neutrino and anti-neutrino have to be indistinguishable. That is, the neutrino would have to be its own antiparticle, as suggested by Majorana.\cite{majorana-orig}

The decay rate for 0$\nu\beta \beta$ can be calculated assuming the exchange of Majorana neutrinos interacting with left handed V-A weak currents.  It is given as
\begin{equation}
\label{decay-rate}
\left( T_{1/2}^{0\nu}\right)^{-1}=G_{0\nu}\left(Q_{\beta \beta},Z\right) \left|M_{o\nu}\right|^2 \langle m_{\beta \beta} \rangle ,
\end{equation}

where $G_{0\nu}\left(Q_{\beta \beta},Z\right)$ is the phase space factor, $\left|M_{o\nu}\right|$ is the nuclear matrix element, and $\langle m_{\beta \beta} \rangle $ is the effective neutrino mass. The effective neutrino mass can be written as a sum over the light neutrino mass states $m_k$, where $U_{ek}$ are the flavor-mass mixing matrix elements:
\begin{equation}
\langle m_{\beta \beta} \rangle = \left| \sum_{k} m_k U_{ek}^2 \right|.
\end{equation}

The phase space factor in Equation \ref{decay-rate} is calculable, and the mixing parameters are constrained by neutrino oscillation measurements.  Thus any measurable rate will give direct information on $\langle m_{\beta \beta} \rangle$ if the nuclear matrix elements are known. Recent theoretical effort has been put into understanding the role of the axial coupling in different nuclei, as well as the impact of the variation in approaches to calculating the nuclear matrix elements. Measurements of $2\nu\beta\beta$, $0\nu\beta\beta$, and excited state decays of heavy nuclei will provide experimental constraints that will complement the theoretical investigations.\cite{EJIRI20191}
\section{\label{sec:MJD} The \textsc{Majorana Demonstrator}}
\begin{figure}[h]
\includegraphics[width=100mm]{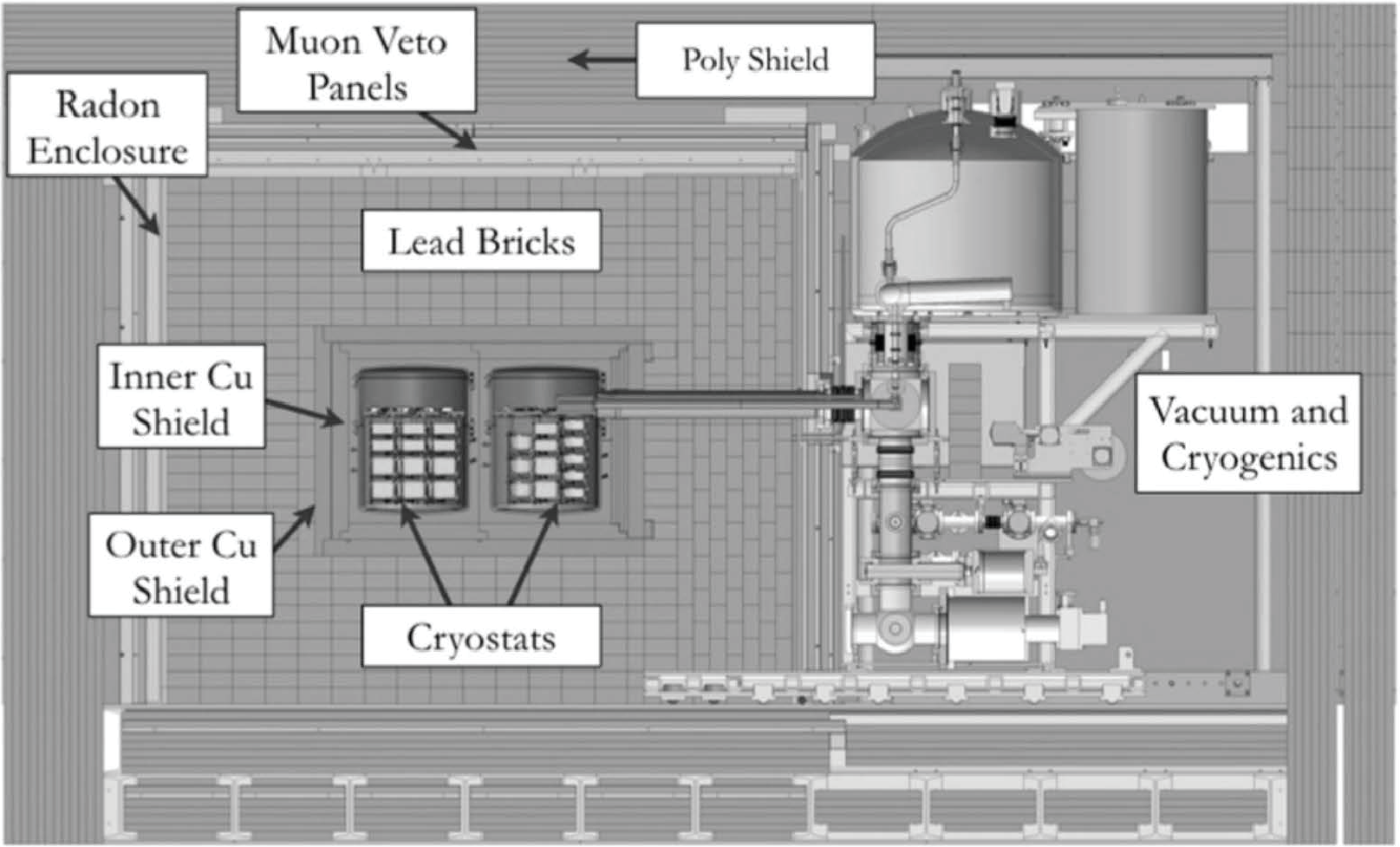}
\caption{\label{fig:MJD-shield} Cut-away view of the \textsc{Demonstrator} shield located on the 4850 level of Sanford Underground Research Facility. The Ge crystals are housed within two copper cryostats at the center of the graded shield.}
\end{figure}

The {\sc Majorana Demonstrator} (MJD) was developed as a critical test of experimental techniques used in the search for neutrinoless double beta decay in germanium via the reaction $^{76}$Ge$\rightarrow ^{76}$Se+e$^-$+e$^-$.  High purity germanium crystals (HPGe) were enriched to a high isotopic fraction of $^{76}$Ge and placed within a graded shield on the 4850 level of the Sanford Underground Research Facility as shown in Figure \ref{fig:MJD-shield}. Special attention was paid to the radiopurity of all experimental parts to keep the background signal as low as possible.  For example, the cryostats and structural components were fabricated from ultralow background copper that was electroformed underground.  The cryostats were surrounded by a lead shield and scintillating panels acting as a veto for cosmic muons. The shield volume was surrounded by 25 cm of polyethelyne to moderate neutrons and the entire volume was purged with nitrogen gas to reduce the radon contribution.   

The \textsc{Demonstrator} accumulated 80 TB of data resulting in 65 kg-yrs of active exposure during the course of operations from 2015-2021. 
The HPGe detector waveforms were recorded by digitizers developed for the GRETINA experiment, with a sampling frequency of 100 MHz and 14 bits of resolution.
 Detector calibration was performed weekly using $^{228}$Th line sources inserted along calibration tracks that wrap around each
cryostat.  The exposure-weighted average
FWHM at 2039 keV across all PPC detectors, including broadening due to gain drift and energy nonlinearities, is 2.52 $\pm$ 0.08 keV.~\cite{Arnquist_2023,PhysRevC.107.045503}
The data were automatically transferred to the National Energy Research Scientific Computing Center (NERSC) for a statistically blind analysis with restricted access to 75\% of the data.
The digitized waveforms from the ``open" 25\% of the data were used to calculate event energies and pulse-shape discrimination (PSD) parameters used to reject likely backgrounds.

\begin{figure}
\includegraphics[width=150mm]{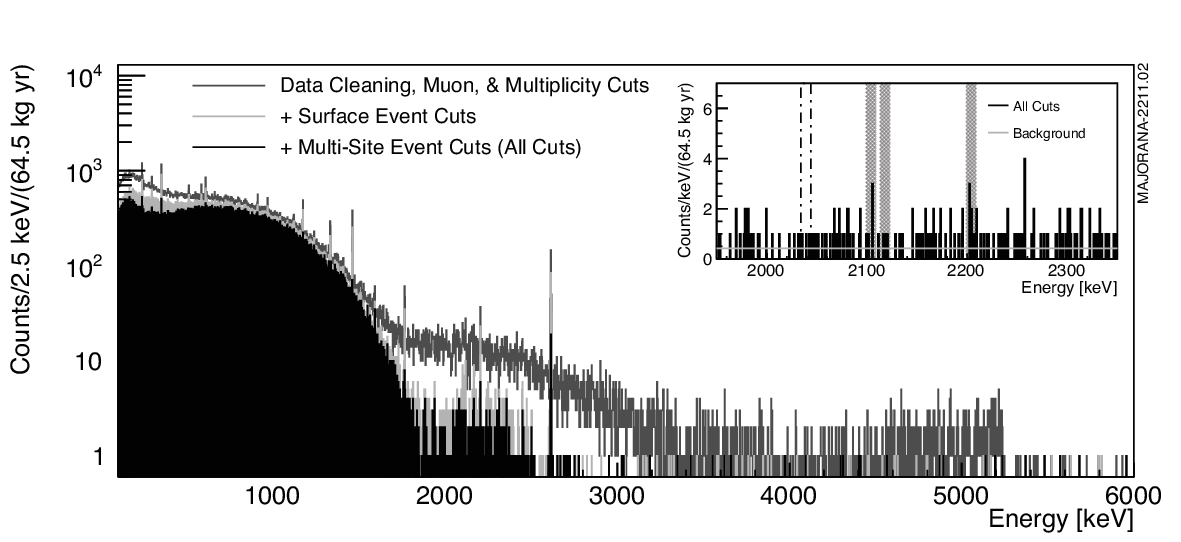}
\caption{\label{fig:MJD-spectrum} The measured energy spectrum above 100 keV for the full enriched exposure after applying multiplicity and data cleaning cuts
(dark gray), DCR, high-AvsE and LQ cuts (light gray), and the low-AvsE cut (black).  Inset contains the measured energy spectrum for the full enriched exposure in the range of 1950–2350 keV, after applying all background cuts. The 360-keV background estimation window excludes the shaded 10 keV windows around expected gamma lines in gray, and the 10 keV window around the 2039~keV Q$\beta\beta$ (dot-dash lines). Horizontal line (gray) is drawn representing the background index. \cite{PhysRevLett.130.062501}}
\end{figure}

Figure \ref{fig:MJD-spectrum} shows the final energy spectrum obtained, with the effect of data cuts visible. 
The rejection of multisite events is achieved via the parameter ``AvsE" which represents a comparison between the maximum amplitude (A) of a waveform current pulse and the total
energy (E) where multisite events have lower A for a given E than single-site events.\cite{PhysRevC.99.065501}
Alpha particles impinging on the passivated surface experience increased charge trapping such that this charge is collected slowly, increasing the slope of the
falling tail of the waveforms relative to bulk events.  The ``Delayed Charge Recovery" (DCR) parameter is a measure of this slope and is used to eliminate surface events.\cite{10.1140/epjc/s10052-022-10161-y} Finally, events with a partial charge deposition in the
transition layer between the n-type surfaces and the detector
bulk experience energy degradation, and have waveforms
with a slow-rise component. We developed the late charge
(LQ) parameter, which is the integral of uncollected charge
after a waveform has reached 80\% of its maximum value to identify these events.
The slow-rise component of these waveforms increases LQ,
and we cut events that fall 5$\sigma$ above the center of the
parameter distribution.\cite{PhysRevLett.130.062501}

The spectrum is
dominated by $2\nu\beta\beta$ below Q$_{\beta\beta}$=2039 keV, and near Q$_{\beta\beta}$ most events are removed by the various cuts. 
Unblinding of data proceeded in a staged fashion with basic data quality assurance checks performed at each stage. 
No changes to analysis parameters or run selection were made after opening the energy window 1950 –
2350~keV which contains Q$_{\beta\beta}$~(inset for Figure \ref{fig:MJD-spectrum} denoted by dot-dash lines). 

A background index was calculated using counts that pass all cuts within a 360-keV
background estimation window, excluding 5 keV around the 2039 keV Q$_{\beta\beta}$ value and expected background gamma
rays at 2103, 2118, and 2204 keV as shown in the inset of Figure \ref{fig:MJD-spectrum}. 
The background rate within this 360-keV window is assumed to be flat. The
surface cuts (DCR, high-AvsE and LQ) remove 85\% of
events in the background region. The multisite cut (lowAvsE) removes 49\% of the remaining events.
After cuts, 153 events remain in the background
estimation window, resulting in a background index of 16.6$^{+1.4}_{-1.3}$ x 10$^{-3}$ cts/(FWHM kg yr).
Upon unblinding the 10 keV window centered on 2039~keV, 4 events were observed, consistent with the background expectation.

From this measurement, a lower limit on the half-life was determined to be T$_{1/2}$ > 8.3 x 10$^{25}$ yr using
\begin{equation}
\label{decay-limit}
T_{1/2} > \text{ln}(2)\frac{NT\epsilon_{tot}}{S},
\end{equation}
where \textit{NT} is the number of $^{76}$Ge nuclei in the active mass, $\epsilon_{tot}$=0.78 is the signal efficiency, and \textit{S} is the upper limit on the signal counts based on the observed data.  
Using the observed half life limit with a range of matrix element values M$_{0\nu}$,  the free nucleon value of g$_A^{eff}$=1.27, and a range of phase space factors, Equation \ref{decay-rate} gives 113 < m$_{\beta\beta}$ < 269 meV.\cite{PhysRevLett.130.062501}

The \textsc{Majorana Demonstrator} has completed it's search for neutrinoless double beta decay where the p-type point contact detectors provided excellent energy resolution.  These detectors are now being combined with the GERDA detector system \cite{PhysRevLett.125.252502} to form the LEGEND experiment with the eventual goal of 1 tonne of enriched material counting for ten years to reach a \textit{T}$_{1/2}$ sensitivity on the order of 10$^{28}$ yr, probing m$_{\beta\beta}$ down to about 10 meV and covering the mass range predicted for the inverted ordering of neutrino masses.\cite{PhysRevC.104.L042501}

\begin{acknowledgments}
This material is supported by a grant (DE-SC0018060) from the U.S. Department of Energy, Office of Science, Office of Nuclear Physics. Support for the \textsc{Majorana Demonstrator} was also provided by the Particle Astrophysics and Nuclear Physics Programs of the National Science Foundation, and the Sanford Underground Research Facility.\end{acknowledgments}

\nocite{*}
\bibliography{tedeschi-mjd}\end{document}